\documentclass[aps,pre,showpacs,preprint]{revtex4}
\usepackage{graphicx}
\usepackage{bm}
\usepackage{amsmath,amssymb}

\begin{document}
\title{Revisit of interfacial free energy of the hard sphere system near hard wall}

\author{ Mingcheng Yang}
\author{Hongru Ma}
\email{hrma@sjtu.edu.cn}

\affiliation{Institute of Theoretical Physics, Shanghai Jiao Tong
University, Shanghai 200240, People's Republic of China}

\begin{abstract}

We propose a simple Monte Carlo method to calculate the interfacial
free energy $\gamma$ between the substrate and the material. Using
this method we investigate the interfacial free energys of the hard
sphere fluid and solid phases near a smooth hard wall. According to
the obtained interfacial free energys of the coexisting fluid and
solid phases and the Young equation we are able to determine the
contact angle with high accuracy, $\cos\theta=1.010(31)$, which
indicates that a smooth hard wall can be wetted completely by the
hard sphere crystal at the interface between the wall and the hard
sphere fluid.

\end{abstract}


\maketitle

\section{Introduction}
The interfacial free energy(IFE) $\gamma$ between different
materials or different phases of a material plays crucial roles in
many physical phenomena such as wetting and spreading \cite{de
Gennes}. An accurate result for $\gamma$ is necessary to understand
these phenomena completely. The hard sphere system as a simple model
system of condensed matter has been investigated extensively. By
changing the density of the system, it undergoes a first order phase
transformation from the liquid phase to the FCC solid phase
\cite{Hoover}.  An interesting problem of the hard sphere system
related to the IFE is whether a smooth hard wall can be wetted
completely by the hard sphere crystal in the density region of
coexistence. The problem was studied by different groups in the past
but a definite answer is still awaiting
\cite{Courtemanche,Courtemanche1, Kegel}. Recent computer simulation
\cite{Dijkstra} by Dijkstra provided a strong evidence of complete
wetting. In that paper the author investigated the effect of  wall
separation on crystalline layers at the walls in a double-wall
system and extracted the complete wetting information from the
phenomenon of surface freezing(including complete wetting and
capillary freezing).  Alternatively, the direct way to study this
wetting behavior is to calculate the cosine of the contact angle in
terms of Young equation, i.e.
$\cos\theta=(\gamma_{fw}^{c}-\gamma_{sw}^{c}) /\gamma_{fs}^{c}$,
here the subscripts $f$, $s$ and $w$ denote respectively the fluid,
solid and wall, and the superscript $c$ denotes the bulk coexisting
phase.  In the complete wetting state $\cos\theta$ given by the
above formula will be larger than unity. The fluid-wall IFE
$\gamma_{fw}$ can be estimated via density functional theory
\cite{Ohnesorge,Gotzelmann}, scaled particle theory \cite{Reiss},
molecular dynamic simulation \cite{Henderson} and Monte Carlo(MC)
simulation \cite{ Heni,Gloor,MacDowell}. However, it is unfeasible
to calculate directly the solid-wall IFE $\gamma_{sw}$ using the
above methods except for the MC simulation \cite{Heni}. Recently,
several groups have calculated the crystal/melt IFE
$\gamma_{fs}^{c}$ of a hard sphere system by different computer
simulation methods \cite{Davidchack, Cacciuto,
Davidchack1,Mu,Davidchack2}. However, the accuracies of the contact
angle $\theta$ from their calculation are too low to conclude
definitely whether or not the wetting is  complete \cite{Dijkstra}.
Comparing with $\gamma_{fs}^{c}$, the error of the contact angle
$\theta$ comes mainly from $\gamma_{fw}^{c}$ and $\gamma_{sw}^{c}$
\cite{Heni,Davidchack2}. Therefore, it is desired to accurately
calculate the IFEs, which provide us not only the confirming of the
complete wetting scenario but also a benchmark to test other
theoretical methods.

In this paper we propose a simple Monte Carlo method of calculating
the IFEs $\gamma_{fw}$ and $\gamma_{sw}$, and obtain the IFEs of the
coexisting hard sphere fluid and solid phase(i.e. $\gamma_{fw}^{c}$
and $\gamma_{sw}^{c}$) with high accuracy, which nearly confirms the
suggestion that at the smooth hard wall-hard sphere fluid interface
a complete wetting phase transition occurs very close to the bulk
freezing transition. Our method is inspired by the cleaving method
of Davidchack and Laird \cite{Davidchack}, where they used two
moving walls to transform the hard sphere system from the bulk phase
into the phase confined completely by two walls and vice versa, and
the IFE can be obtained by calculating the work done in the
reversible process. However, as mentioned in Davidchack's work
\cite{Davidchack}, irreversibility is introduced during cleaving a
fluid system using constructed hard walls. The reason of
irreversibility is that the system doesn't reach the equilibrium
state in a typical run, which can be reduced by increasing the
duration of equilibration run. The problem can be avoided by
designing an auxiliary path which connects the confined and the bulk
phase and the states along the path are all in equilibrium state
during the simulation. The goal is achieved by combine the
 cleaving technique \cite{Davidchack,Broughton} and the so-called
flat histogram methods \cite{Berg,Wang,Wang1}. From these method
$\gamma_{fw}$ and $\gamma_{sw}$ can be accurately estimated.
 In the following sections we employ
Wang-Landau (WL) algorithm \cite{Wang} to demonstrate the
implementation of the current approach.

In order to check the validity of our method, we calculated the IFE
between the hard sphere system and the smooth hard wall at usual
densities, including $\gamma_{fw}$ and $\gamma_{sw}$ for three
different crystal orientations $(1,1,1)$, $(1,1,0)$, and $(1,0,0)$.
The results are consistent with those of Heni and L\"{o}wen within
the statistical error \cite{Heni}. We also investigated the finite
size effect of longitudinal dimension(along $z$ direction) on the
IFE, and then obtained the effective interaction potential between
two hard walls. Finally, we use the method to calculate the IFEs of
the coexisting phase $\gamma_{fw}^{c}$ and $\gamma_{sw}^{c}$, and
then obtained the contact angle $\cos\theta$. The paper is organized
as follows: in section II, the model system and the numerical
algorithm are described. The results are presented in section III.
Section IV is a brief conclusion.

\section{Model and Algorithm}
Our system is composed of $N$ hard sphere particles with diameter
$\sigma$ in a fixed volume $V$. The periodic boundary conditions are
imposed in all three directions. Hard sphere system is athermal,
therefore its thermodynamic properties are solely dependent on the
dimensionless packing fraction $\eta=\pi\rho\sigma^{3}/6$, where
$\rho$ is the number density of the particles. Pairs of particles
$i$ and $j$ interact via the potential
\begin{equation}
U_{Hs}(\vec{r_{i}},\vec{r_{j}})= \left\{
\begin{array}{ll}
\infty, \quad &{\rm when}\quad|\vec{r_{i}}-\vec{r_{j}}|<\sigma,\\
 0, &{\rm otherwise},
\end{array}
\right.
\end{equation}
here $\vec{r}$ is the positions of the particles.

\begin{figure}[htbp]
\includegraphics[width=0.5\textwidth]{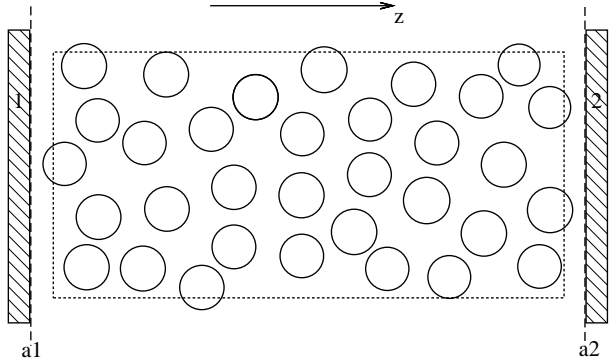}
\caption{Diagram illustrating the change of the system boundary
conditions by two synchronously moving smooth hard walls. When the
walls are respectively placed on $|a_{1}|$ and $|a_{2}|$ ($|a_{1}|=
|a_{2}-Z_{0}|=D_{0}>\sigma/2$), they do not interact with the
particles and the system is in the bulk phase; when the walls are
placed at the boundaries, the system are completely confined by two
hard walls.}\label{fig1}
\end{figure}
Two moving smooth hard walls are used to transform the system from
the periodic boundary conditions into the hard wall boundary
conditions, as Davidchack and Laird did \cite{Davidchack}. The
detail is illustrated in Figure \ref{fig1}. The two walls can be
displaced synchronously along $z$ direction from the boundary of the
system (shown by the dot line in Figure \ref{fig1}) to the position
$a_{1}$ and $a_{2}$ ($|a_{1}|=|a_{2}-Z_{0}|> \sigma/2$, $Z_{0}$ is
the dimension of the system in the $z$-direction), respectively. The
interaction between the walls and the particles is
\begin{equation}
U_{w}(\vec{z_{i}},\vec{z_{w}})= \left\{
\begin{array}{ll}
\infty, \quad & {\rm when }\quad|\vec{z_{i}}-\vec{z_{w}}|<\sigma/2, \\
 0, &{\rm otherwise},
\end{array},
\right.
\end{equation}
where $\vec{z_{i}}$ and $\vec{z_{w}}$ are the coordinates of the
particles and the walls in $z$ direction, respectively. Therefore,
when the separation between the walls and the boundaries is larger
than the hard sphere radius, the walls do not interact with the
particles and the system is in the bulk phase; however, when the
walls are placed at the boundaries, the system are completely
confined by the two hard walls.

The total Hamiltonian of the system $H$ is a sum of the kinetic
energy and the potential energy, given by
\begin{equation}
K=\sum_{i}\frac{P_{i}^{2}}{2m}
\end{equation}
and
\begin{equation}
U=\sum_{i<j}U_{Hs}(\vec{r_{i}},\vec{r_{j}})
+U_{w}(\vec{z_{i}},\vec{z}_{w1},\vec{z}_{w2}),
\end{equation}
where $m$ is the mass of particles and $p_{i}$ is the linear
momenta. The Helmholtz free energy of the system is written as
\begin{equation}  \label{5}
F(T,V,N,z_{w1},z_{w2})=-k_{B}T \;{\rm ln}(Q(T,V,N,z_{w1},z_{w2})),
\end{equation}
where $Q(T,V,N,z_{w1},z_{w2})$ is the canonical partition function
of the system
\begin{equation}
Q(T,V,N,z_{w1},z_{w2})=\frac{1}{\Lambda^{3N}N!}\int
dr_{1}^{3}\cdots\int dr_{N}^{3}\exp(-\beta
U(\{\vec{r}_{i}\},z_{w1},z_{w2})).
\end{equation}
In the leading order approximation, the total IFE is just the
difference of the total free energy between the confined system and
the bulk system \cite{Evans, Heni}. The IFE per unit area $\gamma$
is expressed by
\begin{eqnarray}
\gamma(\eta,T)&=&\frac{F_{confined}(T,\eta)-F_{bulk}(T,\eta)}{A}
\nonumber \\
&=&\frac{F(T,\eta,z_{w1}=0,z_{w2}=Z_{0})-F(T,\eta,z_{w1}=a1,z_{w2}=a2)}{A}, \label{7}
\end{eqnarray}
here $A$ is the total contact area between the system and two hard
walls.

Using WL algorithm \cite{Wang}, initially developed to compute the
density of state(DOS) $g(E)$, we can easily obtain the excess
Helmholtz free energy in a Metropolis-type MC simulation. The
original algorithm does a random walk in energy space with an
acceptance ratio $\min\{1,g(E_{i})/g(E_{f})\}$, where $E_{i}$ and
$E_{f}$ are respectively the energies of the current and the
proposed configuration. The DOS of the accepted energy then adjusted
by multiply it with a factor $f>1$. The histogram is accumulated in
the simulation and its flatness is tested. A new run with reduced
$f$ starts when the histogram reaches a specified flatness
condition. The DOS obtained when the $f$ reduced close to unity.

The same procedure can be used to calculated the excess Helmholtz
free energy. In order to use the method, we employ an expanded
canonical ensemble, where the separation $D_{wb}$ from the hard
walls to the system boundary is an additional ensemble variable. The
expanded ensemble method has been used to calculate the Helmholtz
free energy at a series of temperatures \cite{Lyubartsev} and the
chemical potential of polymers \cite{Wilding} and colloid
\cite{Tej}. Obviously, $0 \leq D_{wb}\leq D_{0}$, where
$D_{0}=|a_{1}| =|a_{2}-Z_{0}|$. We use $D_{wb}$ to rewrite (\ref{7})
as
\begin{eqnarray}
\gamma(\eta,T)=\frac{F(T,\eta,D_{wb}=0)-F(T,\eta,D_{wb}=D_{0})}{A}.    \label{8}
\end{eqnarray}
Every $D_{wb}$ corresponds to a macroscopic state. The DOS
$g(D_{wb})$ of the system can be computed via WL algorithm. For the
hard interaction system, the logarithm of DOS is proportional to the
partition function of the system. According to (\ref{5}) and (\ref{8}) we get
\begin{eqnarray} \label{9}
\gamma(\eta,T)=\frac{k_{B}T\ln(g(D_{wb}=D_{0}))-k_{B}T\ln(g(D_{wb}=0))}{A}.
\end{eqnarray}

The expanded canonical ensemble simulations are performed in a
similar way as in the canonical ensemble. In addition to the
particles translation operation which is accepted or rejected in the
usual way, the trial moves also include the walls displacement along
$z$ axis, which is accepted with the probability
\begin{equation}  \label{10}
P_{acc}\{D_{wb}^{old}\rightarrow D_{wb}^{new}\}=
\min\{1,g(D_{wb}^{old})/g(D_{wb}^{new})\},
\end{equation}
Where $D_{wb}^{old}$ and $D_{wb}^{new}$ are the distances before and
after the hard walls are displaced, respectively. In other words a
random walk is implemented in $D_{wb}$ space instead of energy
space. The calculation is as follows, the interval $[0, D_{0}]$ is
divided into many subintervals, a random walk in this space is
performed in the same way as that of the Wang-Landau method in the
energy space. A initial guess of the DOS is assigned, the walk goes
on with the acceptance probability given by (\ref{9}) and a initial
factor $f>1$, the DOS of the accepted state is modified by multiply
the corresponding DOS with $f$. A histogram of accepted states is
accumulated and the flatness of it is tested. The run is finished
when the histogram is flat enough, here we use the criteria that the
deviation of the histogram away from its average is less than 20\%.
A new run starts with the DOS obtained in the previous run as
initial DOS and the factor $f$ is reduced to $\sqrt{f}$. The full
calculation is ended when the reduce factor $f$ is very close to
$1$, here we use $f-1< 10^{-5} - 10^{-6}$. At the end of the
simulation the $\gamma(\eta,T)$ can be determined from equation
(\ref{9}).

\section{Results and Discussion}

The simulations were performed in a cuboid box with conventional
periodic boundary conditions. The $D_{wb}$ is a continuous variable,
however the WL algorithm demands a discrete $D_{wb}$. In the
calculation the continuous  $D_{wb}$ is replaced with a set of
discrete points and the walls may displace by jumping between
nearest-neighbor points. Increasing $D_{wb}$ and decreasing $D_{wb}$
are attempted with equal frequency, which are accepted or rejected
by transition rule (\ref{10}). At first, we check the validity of
our method by calculating $\gamma_{fw}$ and $\gamma_{sw}$ at two
ordinary densities.

\subsection{Fluid-wall $\gamma$}
\begin{figure}
\includegraphics[angle=0,width=0.45\textwidth]{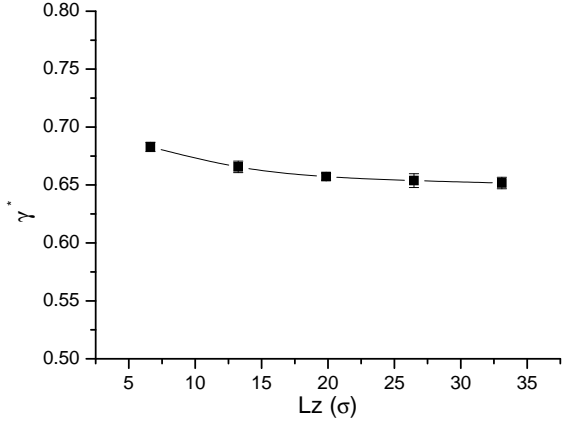}
\caption{Data points (with error bars) show the reduced fluid-wall
interfacial free energy $\gamma_{fw}^{*}$ at $\eta=0.3$. The solid
line is merely a guide to the eye.} \label{fig2}
\end{figure}

In this subsection we report the results of the $\gamma$ for the
fluid-wall interface. The transverse dimensions of the simulation box are respectively
$L_{x}=8.1\sigma$ and $L_{y}=7\sigma$. In order to study the
longitudinal finite size effect we consider several different
longitudinal dimensions $L_{z}=6.6\sigma$, $L_{z}=13.2\sigma$,
$L_{z}=19.8\sigma$, $L_{z}=26.4\sigma$ and $L_{z}=33.1\sigma$, and
the number of particles $N$ are respectively $216$, $432$, $648$, $864$ and
$1080$, which correspond to a packing fraction of $\eta=0.3$. The
dependence of $\gamma_{fw}^{*}=\gamma_{fw}\sigma^{2}$ on $L_{z}$ is
shown in Figure \ref{fig2}. The $\gamma_{fw}^{*}$ decreases
with increasing the longitudinal dimension of the system at fixed
packing fraction and the curve flattens gradually. In other words,
there exists an effective interaction potential $\Phi_{\rm eff}$
between the two walls, which is a function of $L_{z}$
and
can be written as
\begin{equation}
\Phi_{\rm eff}(L_{z})=A(\gamma_{fw}(L_{z})-\gamma_{fw}(\infty)),
\end{equation}
here $A$ is the total contact area and $\gamma_{fw}(\infty)$ is the
exact IFE.

Our result of the $\gamma_{fw}^{*}$ at the largest separation, $L_z=
33.1$,
 is $0.652\pm0.005)kT/\sigma^{2}$, which can be regarded as a good approximation
 of the exact IFE and this value is in
 very good agreement with
that of Heni's \cite{Heni}, $\gamma_{fw}^{*}=(0.656\pm0.03)kT$. We
also performed simulations for a system with $L^{'}_{x} \times
L^{'}_{y} \simeq 2 (L_{x} \times L_{y})$ and $L^{'}_{z}=33.1\sigma$,
the results are basically the same as that of the results obtained from the smaller
system which indicates our system size is already big enough.

\subsection{Solid-wall $\gamma$}
\begin{figure}
\includegraphics[angle=0,width=0.45\textwidth]{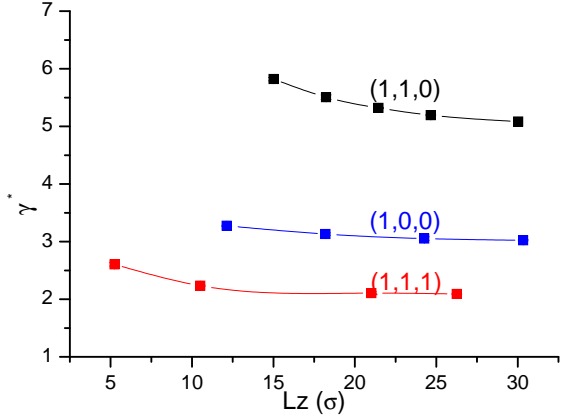}
\caption{Data points show the crystal-wall reduced interfacial free
energy $\gamma_{sw}^{*}$ for three different orientations at
$\eta=0.6$. From bottom to top: (1,1,1), (1,0,0) and (1,1,0)
orientation. The solid line is merely a guide to the eye.
Statistical errors do not exceed the symbol size.}\label{fig3}
\end{figure}

In the solid-wall case the initial configuration is an ideal fcc
lattice with  periodic boundary conditions. The system dimensions
$L_{x}$, $L_{y}$, and $L_{z}$ are given in Table \ref{tab1} and the
number of particles $N$ are determined in such a way that the volume
fraction remains constant, $\eta=0.6$. Here, we calculate the
crystal-wall $\gamma_{sw}$ for the three crystalline orientations
$(1,1,1)$, $(1,1,0)$ and $(1,0,0)$. The results are compared with
those of Heni's in Table \ref{tab1}. The resulting $\gamma_{sw}^{*}$
as a function of $L_{z}$ are plotted in Figure \ref{fig3}. The
profiles of the IFE and  the effective potential are similar to
those of the fluid phase. However, the IFEs are largely anisotropic
and their ratio is $\gamma_{110}:\gamma_{100}:\gamma_{111} \approx
5:3:2$ at $\eta=0.6$, which implies a hard sphere solid prefers to
pack in the $(1,1,1)$ orientation near a smooth hard wall. We also
checked the lateral finite size effects of the crystal-wall
$\gamma_{sw}$ for three orientations, all results are consistent
with those obtained from the smaller systems within statistical
error.
\begin{table}
\begin{tabular}{|c|c|c|c|c|c|c|c|c|c|}
  \hline
      & $L_{x}$  & $L_{y}$  & $L_{z1}$  & $L_{z2}$  & $L_{z3}$
      &$L_{z4}$  & $L_{z5}$ & $\gamma^{*}$ &$\gamma^{*}_{1}$\\\hline
  111 &6.4$\sigma$&5.6$\sigma$&5.3$\sigma$&10.6$\sigma$&21$\sigma$&26.3$\sigma$&&2.09$\pm$0.01&1.74$\pm$0.21\\\hline
  100 &6.1$\sigma$&6.1$\sigma$&12.1$\sigma$&18.2$\sigma$&24.1$\sigma$&30.3$\sigma$&&3.02$\pm$0.01&2.95$\pm$0.30\\\hline
  110 &6.1$\sigma$&8.6$\sigma$&15$\sigma$&18.2$\sigma$&21.4$\sigma$&24.7$\sigma$&30$\sigma$&5.08$\pm$0.01&5.03$\pm$0.30\\
  \hline
\end{tabular}
\caption{The system dimensions and interfacial free energys with
their statistical error for three crystal face orientations.
$\gamma^{*}$: our simulation result; $\gamma^{*}_{1}$: MC simulation
form Ref. [11].}\label{tab1}
\end{table}

\subsection{Wetting}

\begin{figure}
\includegraphics[angle=0,width=0.45\textwidth]{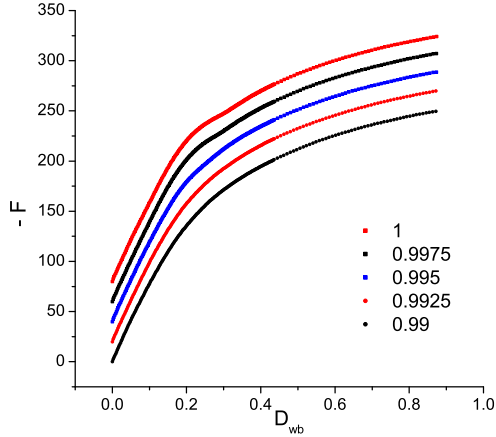}
\caption{(color online)The relative Helmholtz free energy versus the
parameter $D_{wb}$ for different packing fractions. From top to
bottom the packing fraction $\eta=\eta_{f}$, $0.9975\eta_{f}$,
$0.995\eta_{f}$, $0.9925\eta_{f}$ and $0.99\eta_{f}$. The shape of
the upper curves obviously differs from the lower curves, which
indicates the formation of the crystalline layers. The curves are
shifted vertically for clarity.}\label{fig4}
\end{figure}

\begin{figure}
\includegraphics[angle=0,width=0.45\textwidth]{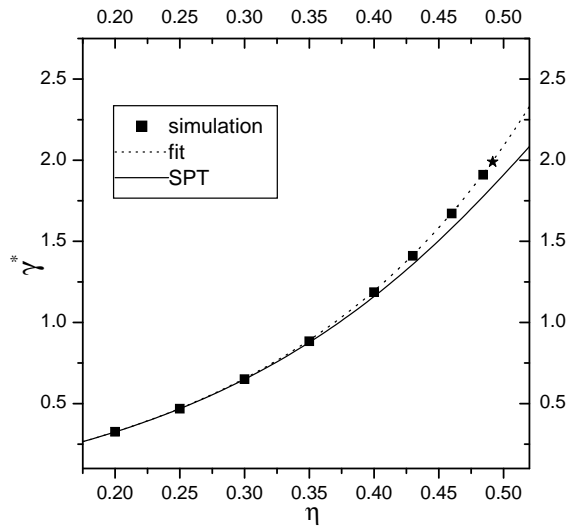}
\caption{The fluid-wall reduced surface tension versus the volume
fraction. The solid line denotes SPT; the dashed line denotes a
numerical fit from SPT; the filled square refers to our simulation
results and $\star$ to the $\gamma_{fw}^{c*}$ obtained from
extrapolation.}    \label{fig5}
\end{figure}

The IFEs of the coexisting phases $\gamma_{fw}^{c}$ and
$\gamma_{sw}^{c}$ are calculated in this section. Based on the
results of the last subsections, we expect that the simulation
systems with lateral dimensions $L_{x}\times
L_{y}\simeq60\sigma^{2}$ and longitudinal dimension
$L_{z}\simeq38\sigma$ is large enough to effectively reduce the
finite-size effect. It is worth emphasizing that the coexisting
volume fractions of the hard sphere fluid and solid phases are
$\eta_{f}=0.4917$ and $\eta_{s}=0.543$ \cite{Davidchack,
Mu,Davidchack2,Davidchack3}, respectively, which are slightly
different from those obtained by Hoover and Ree \cite{Hoover}.
Because the hard sphere particles usually crystalize on a smooth
hard wall with the $(1,1,1)$ orientation, we need only consider this
packing structure. The $\gamma_{sw}^{c}$ can be easily determined,
however, the prefreezing phenomenon prevents us from directly
achieving the $\gamma_{fw}^{c}$ from the simulation. We have to
extrapolate the $\gamma_{fw}$ from lower packing fraction to the
bulk coexisting packing fraction $\eta_{f}$ \cite{Heni}. In order to
get a more accurate estimation of the $\gamma_{fw}$ at the bulk
coexisting packing fraction, we should use the data at packing
fractions as close to the bulk coexisting as possible, on the  other
hand, the data used should correspond to fluid phases, i.e., before
the prefreezing occurs. Therefore, it's crucial to accurately locate
the volume fraction just before the prefreezing. Our simulation
results are the relative Helmholtz free energy of the system with
various $D_{wb}$, therefore, if the prefreezing occurs so that
crystalline layers appear at some  $D_{wb}$ value, a singularity may
be observed on the free energy versus $D_{wb}$ curve.
  Figure \ref{fig4}  depicts the free
energy-$D_{wb}$ curve for different packing fractions $\eta$ close
to the bulk coexisting packing. The volume fractions of the system
from top to bottom are respectively $\eta_{f}$, $0.9975\eta_{f}$,
$0.995\eta_{f}$, $0.9925\eta_{f}$ and $0.99\eta_{f}$. From Figure
\ref{fig4} we see that the packing fraction of forming the
crystalline layers is lower than $0.995\eta_{f}$, however it is
difficult to locate the point exactly. With these observations it is
appropriate to extrapolate from the volume fraction
$\eta=0.99\eta_{f}$. On the other hand,  by observing the density
profile near the walls Dijkstra \cite{Dijkstra} suggested that at
the packing fraction $\eta\sim0.987\eta_{f}$ the crystalline layers
begin to form. To check the consistence of the extrapolation, we
also obtained the IFE by extrapolations from $\gamma_{fw}$ at the
volume fraction $\eta=0.985\eta_{f}=0.4843$. The two extrapolation
results are the same within statistical error of the simulation. The
results of the IFE are shown in Table \ref{tab2}. In Figure
\ref{fig5} we compare our simulation results with those from the
scaled particle theory combined with the Carnahan-Starling equation
of state(SPT) \cite{Heni}. Here we don't present a comparison with
Heni's simulation results, as the coexistence densities taken in
their simulation are different from ours.

\begin{table}[!h]
\tabcolsep 0pt \vspace*{-12pt}
\begin{center}
\def\temptablewidth{0.5\textwidth}
{\rule{\temptablewidth}{1pt}}
\begin{tabular*}{\temptablewidth}{@{\extracolsep{\fill}}ccc}
$\eta$ & $\gamma^{*}$ &  $\delta$ \\
\hline
 0.20  &  0.3261   & 0.0004 \\
 0.25  &  0.4685   & 0.0004 \\
 0.30  &  0.6499   & 0.0004 \\
 0.35  &  0.8836   & 0.0011 \\
 0.40  &  1.1866   & 0.0007 \\
 0.43  &  1.4106   & 0.0010 \\
 0.46  &  1.6711   & 0.0011 \\
 0.4843 &  1.9105   & 0.0016 \\
 0.4867 & 1.9323    & 0.0018 \\
 0.4917$_{f}$ &  1.9872  & 0.0012 \\
 0.4917$_{f_1}$ &  1.9896  & 0.0016 \\
 0.5430$_{s}$&  1.4380   & 0.0050 \\
\end{tabular*}
{\rule{\temptablewidth}{1pt}}
\end{center}
\caption{The reduced surface free energys $\gamma^{*}$ with its
error $\delta$ for different volume fractions. The subscript $f$ and
$f_{1}$ refer to the surface tension of coexisting hard sphere fluid
phase $\gamma_{fw}^{c*}$ obtained by extrapolating from
$0.99\eta_{f}$ and $0.985\eta_{f}$, respectively. And the subscript $s$ refers to the
surface free energy of coexisting hard sphere solid phase
$\gamma_{sw}^{c*}$ with $(1,1,1)$ face. }\label{tab2}
\end{table}

Using the available latest crystal/melt IFE $\gamma_{fs}=0.546(16)$
from simulation \cite{Davidchack2} combined with our results, the
contact angle $\theta$ can be calculated in terms of the Young
equation, $\cos\theta=(\gamma_{fw}^{c}- \gamma_{sw}^{c})
/\gamma_{fs}^{c}=1.010(31)$, which nearly indicates that a complete
wetting phase transition will occur. If the above $\gamma_{fs}^{c}$
is replaced by the value $0.55(2)$ obtained from nucleation
experiment \cite{Marr}, which is the average of the crystal/melt
IFEs over all three orientations and larger than the
$\gamma_{fs}^{c}$ for (1,1,1) interface, we will have the same
conclusion. The main source of error is the crystal/melt IFE
$\gamma_{fs}^{c}$, a more accurate $\gamma_{fs}^{c}$ is necessary to
draw a definite conclusion. Of course, one can also improve the
precision of the $\gamma_{fw}^{c}$ and $\gamma_{sw}^{c}$ by
employing other sampling method, e.g. multicanonical algrithm
\cite{Berg}.

\section{conclusion}

In conclusion, we proposed a simple Monte Carlo method and
calculated the IFEs of the hard sphere system at the smooth hard
wall with high accuracy. The contact angle calculated with the
accurate IFEs $\gamma_{fw}^{c}$ and $\gamma_{sw}^{c}$ make us
confident that a smooth hard wall can be completely wetted by the
hard sphere crystal at the wall-hard sphere fluid interface. Our
conclusion is consistent with Dijkstra's work. However, when using
the Young equation to study the wetting behavior, the finite-size
effect can not be avoided completely as Dijkstra did in Ref. [6].
Therefore larger system and higher precision are necessary to
confirm unambiguously the complete wetting phenomenon. Furthermore,
the IFEs obtained also provide a benchmark to test other theoretical
approaches. Our method can be extended to other systems (e.g. soft
sphere system, polydisperse hard sphere system and hard/soft wall
system) in a straightforward manner. We have already obtained  some
results of the polydisperse hard sphere system which will be
reported elsewhere.


\emph{Note added in proof}. --- After the completion of this study,
we are aware of Laird et al studied the same problem with molecular
dynamics simulation and have the same conclusions as ours
\cite{Laird}. The work is supported by the National Natural Science
Foundation of China under grant No.10334020 and in part by the
National Minister of Education Program for Changjiang Scholars and
Innovative Research Team in University.

\end{document}